# Vortex motion in reconfigurable three-dimensional superconducting nanoarchitectures

Authors: Elina Zhakina,*[1] Luke Turnbull,[1] Weijie Xu,[1] Markus König,[1] Paul Simon,[1] Wilder Carrillo-Cabrera,[1] Amalio Fernandez-Pacheco,[2] Uri Vool[1], Dieter Suess[3,4], Claas Abert[3,4], Vladimir M. Fomin,[5, 6] and Claire Donnelly*[1,8]

[1]Max Planck Institute for Chemical Physics of Solids, Nöthnitzer Str. 40, 01187 Dresden
[2] Institute of Applied Physics, TU Wien, Wiedner Hauptstr. 8-10/134,1040 Vienna, Austria
[3] Physics of Functional Materials, Faculty of Physics, University of Vienna, Kolingasse 14-16, A-1090, Vienna, Austria
[4]Research Platform MMM Mathematics-Magnetism Materials, University of Vienna, Vienna, 1090, Austria
[6]Institute for Emerging Electronic Technologies, Leibniz IFW Dresden, Helmholtzstraße 20, D-01069 Dresden, Germany
[7] Faculty of Physics and Engineering, Moldova State University, strada A. Mateevici 60, MD-2009 Chişinău, Republic of Moldova
[8]International Institute for Sustainability with Knotted Chiral Meta Matter (WPI-SKCM2), Hiroshima University, Hiroshima 739-8526, Japan

*Correspondence to: elina.zhakina@cpfs.mpg.de, claire.donnelly@cpfs.mpg.de

**Abstract:**

When materials are patterned in three dimensions, there exist opportunities to tailor and create functionalities associated with an increase in complexity, the breaking of symmetries, and the introduction of curvature and non-trivial topologies[1–6]. For superconducting nanostructures, the extension to the third dimension may trigger the emergence of new physical phenomena [7–9], as well as advances in technologies. Here, we harness three-dimensional (3D) nanopatterning to fabricate and control the emergent properties of a 3D superconducting nanostructure. Not only are we able to demonstrate the existence and motion of superconducting vortices in 3D but, with simulations, we show that the confinement leads to a well-defined bending of the vortices within the volume of the structure. Moreover, we experimentally observe a strong geometrical anisotropy of the critical field, through which we achieve the reconfigurable coexistence of superconducting and normal states in our 3D superconducting architecture, and the local definition of weak links. In this way, we uncover an intermediate regime of nanosuperconductivity, where the vortex state is truly three-dimensional and can be designed and manipulated by geometrical confinement. This insight into the influence of 3D geometries on superconducting properties offers a route to local reconfigurable control for future computing devices, sensors, and quantum technologies.

**Main Text:**

The dimensionality of a system determines its functionality. While planar devices dominate many fields, the extension to three dimensions offers an opportunity both to overcome fundamental limitations and achieve new functionalities. In semiconductors, for example, limitations in miniaturisation have meant that two-dimensional 2D devices no longer follow Moore's law: the move to 3D stacked CMOS offers a route to increase both device density and interconnectivity. In optics, the design of 3D metamaterials offers new control over the properties of light, from broadband polarisation[3] to negative refractive indices[4]. In magnetism, the move to 3D not only results in new dynamic properties and topological solitons [1,10]: properties generally achieved via the design of crystals or interfaces can be mimicked by patterning simple materials in 3D geometries.

With the move to three dimensions offering progress in a number of areas of research, there is likewise a multitude of opportunities for 3D nanosuperconductors. Indeed, from a fundamental point of view, the introduction of three dimensionality offers local control over superconducting vortices [7–9], while the introduction of 3D topologically non-trivial geometries have been predicted to lead to new quantum phenomena [11–16] such as a so-called 'nodal state' in a superconducting Möbius strip [17]. As well as offering opportunities for fundamental research, the increase in dimensionality leads to opportunities to advance technologies, such as the extension of superconducting sensors with ultra-sensitive detection of magnetic fields[18,19] as well as single photon detectors[20] to 3D spatial and component resolution. In addition, superconductor-based computing including energy-efficient neuromorphics[21,22,23] and quantum computing[24,25] will benefit from the increased interconnectivity and complexity offered by 3D geometries.

Amidst growing interest in 3D nanosuperconductivity, there have been a number of developments in nanofabrication, combining both ion-beam deposition of superconducting nanowires [26,27], as well as strain-induced rolling of thin films into 3D space [28–32]. In this way, first experimental observations of non-trivial effects have been achieved such as geometry-induced phase slips in nanohelices, and improvements in microwave radiation detection in 3D rolled helices [28]. However, while these first works pave a route to the realisation of 3D superconducting nanostructures, the full potential of 3D nanosuperconducting architectures – from the physics of 3D dynamics and geometric effects, to the realisation of 3D devices – remains to be explored.

Here, we study the effect of the 3D confinement geometry on a superconductor by harnessing 3D nanopatterning to fabricate and geometrically control the emergent properties of a 3D superconducting nanostructure. With transport measurements, we determine the superconducting properties and confirm the presence and propagation of superconducting vortices through our 3D superconducting conduit. Through the development of a finite-element three-dimensional time-dependent Ginsburg-Landau (3D TDGL) model, we determine the superconducting vortices to be curved. As well as the curved nature of the vortices, our experiments show how the geometric confinement results in an effective anisotropy in the critical field, that provides local, reconfigurable control over the superconducting state and the controlled introduction of weak links, providing new possibilities in the design and control for prospective applications [33].

To establish the fabrication of 3D superconducting nanostructures with arbitrary geometries, we harness 3D focused electron beam induced deposition (3DFEBID, shown in Fig. 1a), a direct-write technique with which it is possible to pattern 3D nanostructures of complex geometry[34–37]. Until now, this technique has not been exploited for 3D superconducting nanostructures. However, in the field of 3D nanomagnetism, the patterning of 3D magnetic nanostructures with FEBID has been critical to experimental developments [38]. Here we pattern a superconducting tungsten-carbide (W-C) 3D nanostructure directly onto electrical contacts with FEBID (see Methods), shown in Fig. 1b. The structure is a prototype of a 3D nanowire network consisting of four base legs grown onto electrical contacts, that are connected by two nanowires meeting at a vertex of approximately 60°, with a cylindrical cross-section of radius 100 nm. This circuit provides an in-built 4-probe measurement of the top vertex[39], which we harness as a conduit for superconducting vortex motion in 3D, as well as to explore 3D geometric anisotropy. The electrical transport measurements confirm that the structure exhibits a sharp superconducting (SC) transition at around 5 K, as shown in Fig. 1c.

To characterise the superconducting properties of the deposited structure, we map the upper critical magnetic field $\mu_0 H_{c2}(T)$ as a function of temperature (Fig. 1d), which provides insight into the effective dimensionality of a system. For our 3D nanostructure, we find that the critical field follows a power law associated with type II superconductors, that allows us to identify an intermediate regime of superconductivity (see Methods) between the two limiting cases of bulk and one dimensional (1D) that occurs due to the nanoscale confinement. This intermediate regime occurs due to the length scale on which we pattern the nanostructure: the diameter of the nanowire building blocks of the 3D nanostructure lies in between the Cooper pair coherence length $\xi$ and the London penetration depth $\lambda$ (see Fig.1e and Methods)[27]. In this intermediate regime, the geometry of the nanostructure is crucial to its superconducting behaviour.

Having confirmed that our 3D nanostructure exhibits type II superconductivity, we next consider the vortex state, as well as vortex motion, in 3D – a key aspect for possible fluxonic applications. In particular, we perform non-local transport measurements [40–42] in which vortices are propagated by a local Lorentz force $F_L$ (see Fig. 2a) that is applied at one end of the 3D conduit (i). In point i of the nanostructure, the applied electrical current ($I$) is locally aligned along the $x$-direction which, in combination with $\boldsymbol{B}$ applied along the $z$-direction, gives the Lorentz force $\boldsymbol{F_L} \sim \boldsymbol{I} \times \boldsymbol{B}$ along the $y$-direction. For a positive current, the component of the Lorentz force along the conduit $\sim F_L*\sin(\beta)$ (where $\beta$ is determined by the geometry in Fig. 2a) acts to "push" the vortices at point i. As the conduit is populated by vortices, this local propagation leads to a long-range transfer of momentum via the repulsive vortex-vortex interaction, and thus the motion of vortices in 3D. The non-local motion of vortices will result in a non-zero electric field associated with this motion, that will generate a potential difference at the other end of the conduit at point ii ($V_{nl}$) in the voltage lead (blue) [40,42,43].

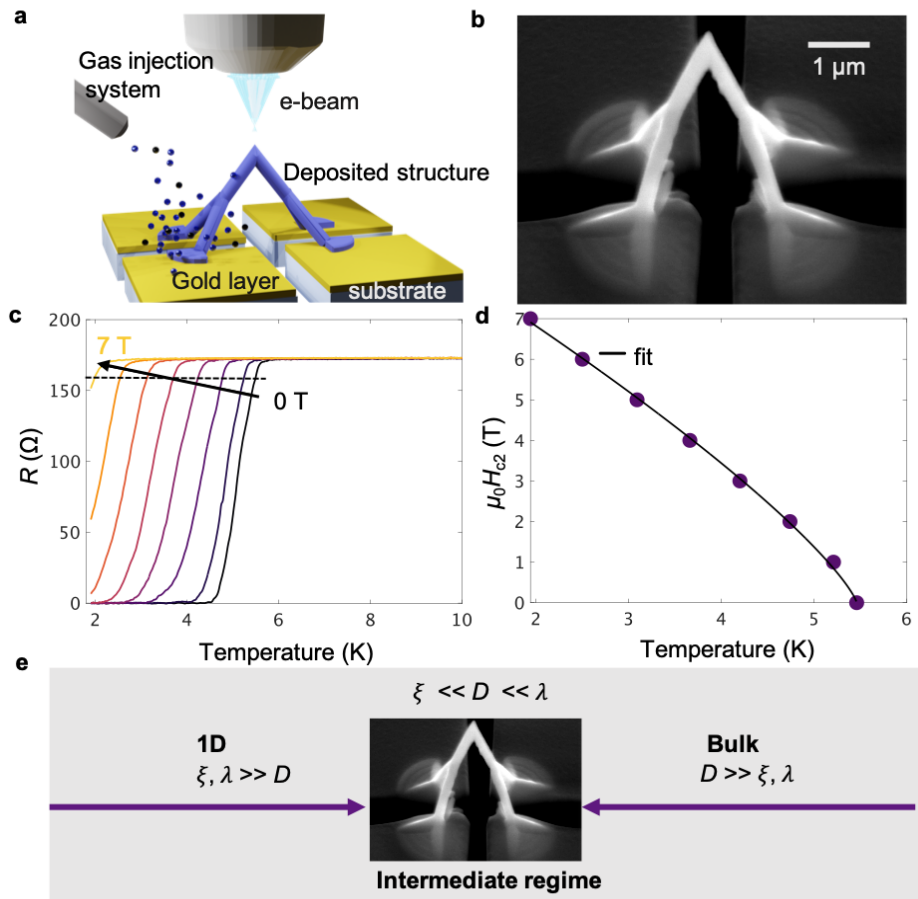

**Figure 1. Fabrication and characterisation of three-dimensional superconducting nanostructures.** (a) Schematic of the FEBID of 3D nanoarchitectures. (b) SEM image of the W-C nanobridge deposited using FEBID. The diameter $D$ of the nanobridge leg is approximately 200 ± 15 nm. (c) Temperature dependence of the resistance of the nanobridge under a magnetic field perpendicular to the substrate, from 0 T to 7 T, measured with $I$ = 100 nA. The dashed line indicates the upper critical magnetic field. (d) Upper critical magnetic field as a function of temperature. Data is fitted to a power-law equation $\mu_0 H_{c2}(T) \sim (1 - T/T_c)^n$ associated with Type II superconductivity.

We perform this non-local measurement to confirm the presence and motion of vortices experimentally. For this, we sweep the magnetic field $B$ from -9 T to 9 T and measure the non-local magnetoresistance $R_{nl} = V_{nl}/I$ in Fig. 2b via low-frequency first harmonic lock-in measurements for a current ($I$) of magnitude 5 µA for temperatures above (6 K, orange line) and below (2 K, black line) the critical temperature (see Methods). At 2 K, two pronounced peaks symmetric with respect to B = 0 are observed in $R_{nl}$, resulting from the potential difference driven by the motion of the vortices under the applied Lorentz force (Fig. 2a). These peaks emerge below the critical temperature and persist deep into the superconducting state, confirming the propagation of vortices through the 3D superconducting nanostructure. Above the critical temperature, there is no peak in the non-local resistance, as expected for the normal state with no superconducting vortices.

In type II superconductors, defects, impurities, or irregularities within the superconducting material can act as pinning centres, hindering the movement of vortices [44]. We gain insight into the origin of the pinning of the vortices in our FEBID WC nanostructure by determining the depinning current $I_d$ as a function of the field in Fig. 2c, indicated by stars. When we calculate the pinning force $F_p$ determined from the applied Lorentz force $F_p = F_L \sim I_d B_z$ (Fig. 2d), we observe a non-linear field dependence with a maximum pinning force for the field $\approx 0.5\ H_{c2}$, that is consistent with volume-dominated pinning [45].

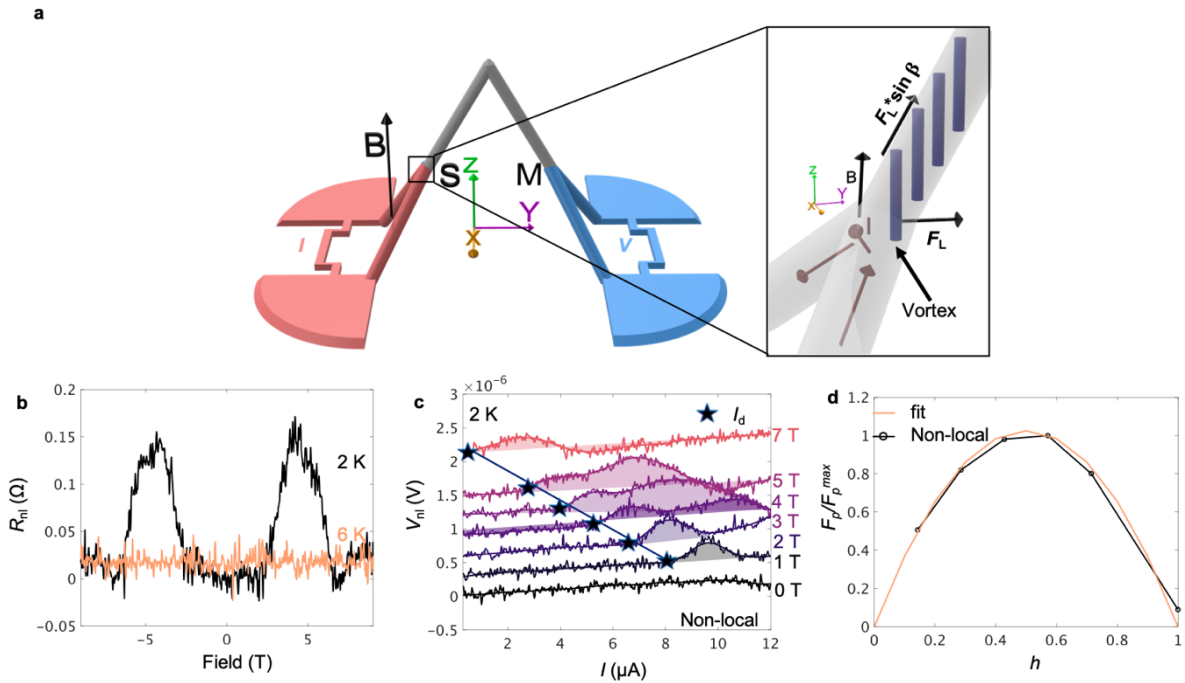

**Figure 2. Propagation of superconducting vortices through the 3D nanostructure.** (a) Schematic of the 4-point electrical transport measurements in a non-local geometry, where $\beta$ is an angle between the long axis of the leg of the nanostructure and the magnetic field. $\beta \sim 30°$ for this nanostructure. (b) Magnetoresistance of the nanostructure in the non-local geometry at temperatures of 2 K and 6 K under a perpendicular magnetic field. $I = 5\ \mu A$. (c) Voltage-vs-Current characteristics of the nanostructure in a non-local geometry under a magnetic field perpendicular to the substrate, from 0 T to 7 T. $I_d$ is the depinning current, indicated by stars. (d) Calculated pinning force as a function of the reduced field $h = H/H_{c2}$ and fitted to the model $\sim h(1-h)$, for volume pinning [45].

This evidence of volume-dominated pinning means that it is important to consider the three-dimensional structure of the vortices and the effect of the 3D confinement. Indeed, in contrast to the flat pancake model, where vortices are approximated as 2D, or straight tubes, in reality vortices can exhibit more complex structures, with the possibility to entangle, and even cut one another [46]. To determine the influence of the 3D geometrical confinement on the 3D structure of the vortices, we perform three-dimensional finite element time-dependent Ginsburg-Landau (3D TDGL) simulations (see Methods). For this, we use a cylindrical nanowire of radius 100 nm, as the building block of our 3D nanobridge and the value of $\kappa = \lambda/\xi$ extracted from the experimental data in Fig. 1 c and d (see Methods). We begin with the simplest case, when the magnetic field is applied perpendicular to the long axis of the nanowire, and plot the Cooper pair density in the horizontal cross-section of the nanowire, both for zero field values (Fig 3a) and finite field (Fig 3b), revealing the formation of superconducting vortices. When the 3D structure of the vortices within the volume of the wire is considered in Fig. 3c, a strong influence of the 3D confinement is observed: the vortices exhibit a curved, 3D structure due to the geometrically enhanced surface effects [47–49]. This effect is more pronounced when the field is applied at an oblique angle to the long axis of the nanowire, as seen in Fig. 3f.

The large surface-to-volume ratio in the confined system leads to a field-dependent competition between the vortex aligning along the magnetic field and meeting the surface of the nanowire at 90°. Higher fields thus lead to a straightening of the vortices, as shown in Fig. 3d, e. This field-dependent curvature indicates the "softness" of the vortices, which has been predicted to strongly influence their pinning, with softer, more deformable vortices more susceptible to pinning and stiffer vortices freer to move [50,51]. Remarkably, this behaviour is reminiscent of the classical model of cutting a soft solid with a flexible wire [52], where the pinning is dominated by volume pinning: here the field-dependent softness of our vortices is consistent with the volume-dominated pinning observed in our structure.

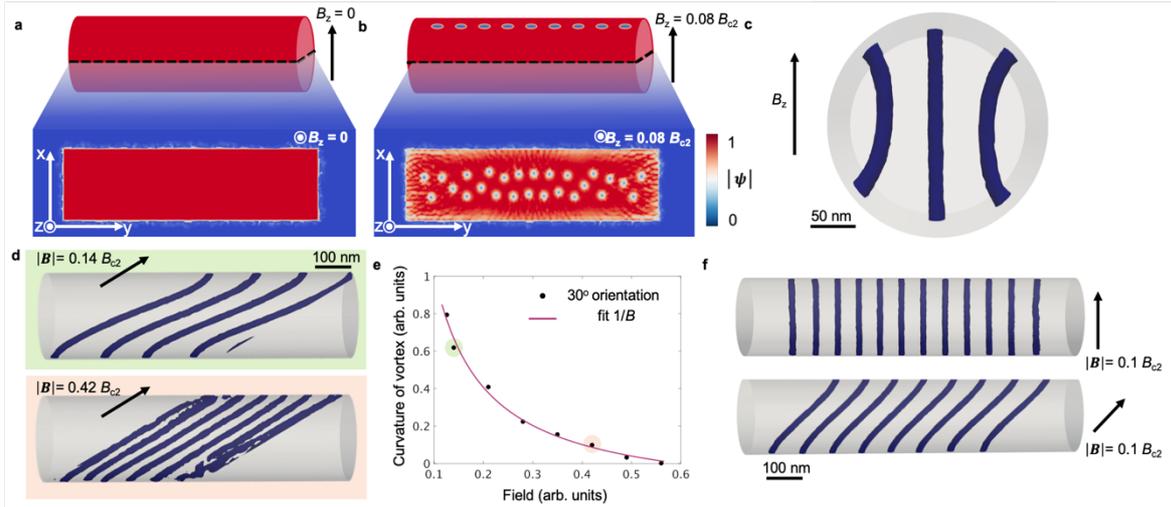

Figure 3. **Finite element three-dimensional time-dependent Ginsburg-Landau (3D TDGL) simulations of a cylindrical nanowire of radius 100 nm**. Simulations shown for a magnetic field applied perpendicular to its long axis (a) for zero magnetic and (b) finite magnetic fields. (c) Cross section of the nanowire with the magnetic field applied perpendicular to its long axis. (d) Side view of the nanowire with two values of magnetic field applied at an angle. (e) Curvature of the vortex as a function of magnetic field. (f) Side view of the nanowire demonstrating the influence of the angle of the applied magnetic field on curvature of the vortex.

By controlling the shape of the superconducting vortices, the geometric confinement induces an anisotropy in the superconducting properties. We probe this anisotropy by measuring the 3D angular dependence of the upper critical field $H_{c2}$ in our 3D SC nanobridge. Specifically, we perform magnetoresistance measurements varying the angle of the applied field with respect to the plane of the 3D nanostructure. For magnetoresistance measurements, the current passes through the nanowire vertex, as shown in Fig. 4a (see Methods). We first rotate the field about the $y$-axis, shown in Figure 4a. The magnetoresistance and its derivative with respect to the field are given in Fig. 4b and inset, respectively, where one can observe a strong angular dependence of the magnetotransport signal. To highlight the superconducting to normal transition, and therefore the field range corresponding to the presence of vortices, we plot the derivative as a function of the angle in a bivariate histogram in Fig. 4c. Here, one can observe a strong angular dependence of the critical field, that varies by approximately 25%, or 1 T, and which is minimal when the field is oriented at 90° to the plane of the nanowire vertex. As the resistance at a normal state in Fig. 4b does not exhibit any angular dependence the 3D nanostructure indicating the isotropic nature confirmed by TEM studies showing isotropic nanocrystallinity on the nm lengthscale (see Methods), we attribute this anisotropy to the geometry of the nanostructure.

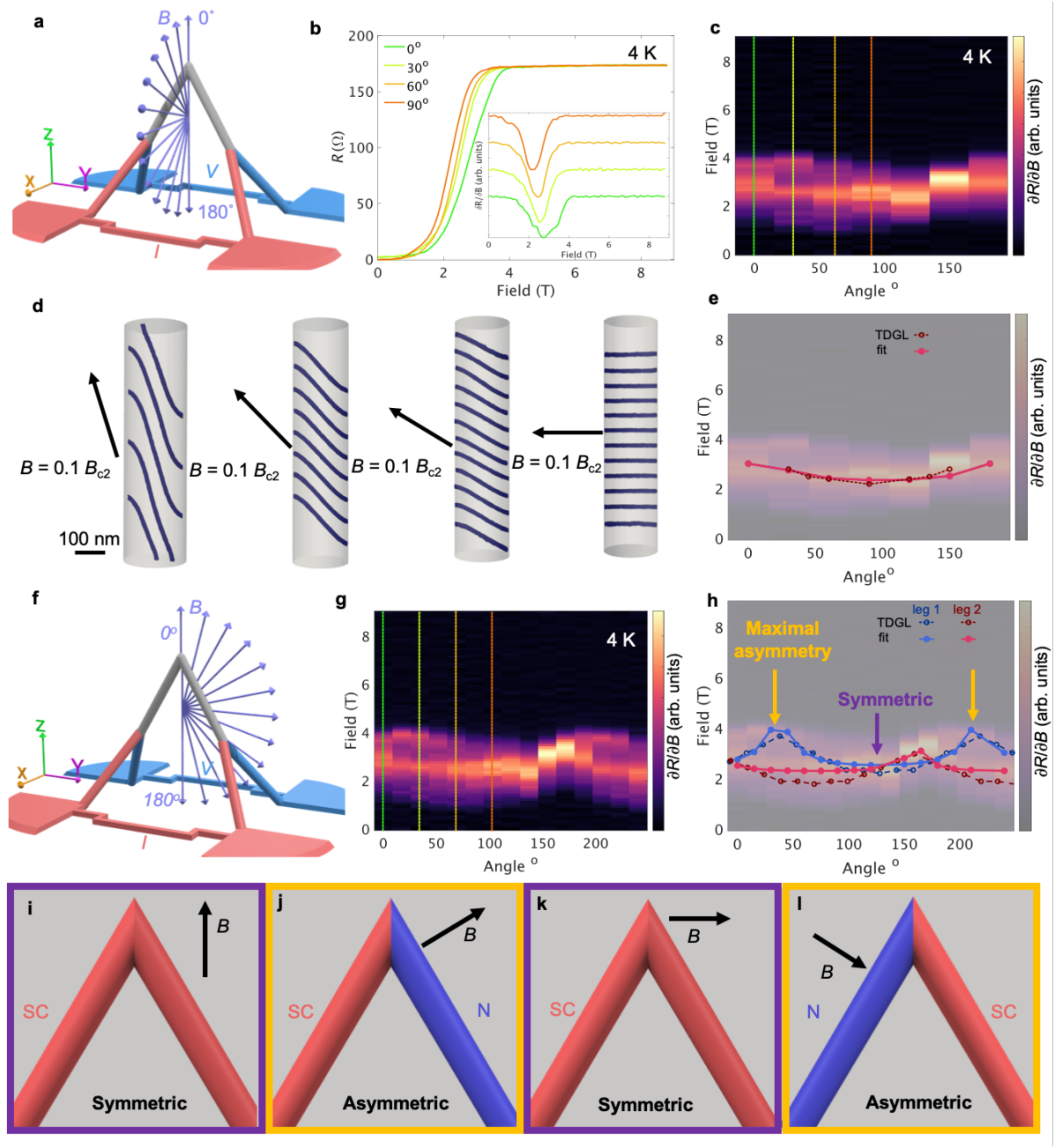

**Figure 4. Geometric anisotropy of the superconducting critical field in the 3D nanostructure.** Sketch of the magnetic field rotation (a) in the *x-z* plane and (f) in the *y-z* plane. (b) Magnetoresistance at 4 K at different magnetic field orientations for a rotation in the *x-z* plane. Insert: derivative of the magnetoresistance with respect to the field. Color bar plot of the derivative of the magnetoresistance for rotation of the magnetic field (c) in the *x-z* plane and (g) in the *y-z* plane, showing the angular dependence of the critical field. (d) 3D TDGL simulation showing the change in the vortex curvature with the change of the magnetic field orientation. The upper critical magnetic field extracted from the 3D TDGL simulation of the nanobridge and the modelled angular dependent upper critical magnetic field, both plotted on top of experimental data for the rotation of the magnetic field I in the *x-z* plane (e) and (h) *y-z* plane. (i-l) sketch showing the reconfigurable coexistence of SC and normal states in the nanostructure.

This geometric anisotropy is a result of the confinement of the nanostructure: at the surface, the vortices meet the boundary at 90° due to the continuity of the supercurrents, while in the volume of the nanowire, the vortices align parallel to the external field [47–49]. We model this anisotropy with a correction to the critical field proportional to the magnetic field component orthogonal to the structure's surface [53] (see Methods) and plot it on top of experimental data in Fig. 4e. The model reproduces the observed anisotropy of the nanostructure, experimentally demonstrating the surface-induced geometric anisotropy in the 3D SC nanobridge. To determine the underlying mechanism for this anisotropy, we next simulate the shape of the vortices as a function of the angle between the long axis of a nanowire and the applied magnetic field (Fig. 4d), using 3D TDGL simulations. We observe a strong angular dependence of the 3D shape of the vortices, with the curvature of the vortices increasing as the angle of the field to the nanowire axis is decreased from 90°. When we apply these simulations to the geometry of our 3D nanostructure – based on our measured $\xi$ and $\lambda$, and no other fitting – we observe a remarkable agreement with the experimental data, both reproducing the qualitative angular dependence of the magnetoresistance, as well as the relative decrease of the critical magnetic field of almost 30 %, as shown in the insert of Fig. 4e.

This 3D anisotropy offers the possibility to locally control the superconducting state in different regions of a 3D nanostructure. To explore this in our nanobridge, we rotate the magnetic field in the *y-z* plane, so that the legs of the conduit are oriented at different angles to the magnetic field. We immediately observe a more complex angular dependence of the critical field (Fig. 4g), with a splitting of the superconducting to normal metal (SC-N) transition indicated by two peaks forming in the derivative. The largest splitting of the magnetoresistance is observed for the magnetic field oriented at around 30°, which corresponds to the maximal asymmetry in the orientation of the magnetic field for the two legs. This split transition indicates that the two legs of the conduit undergo a SC-N transition at different magnetic fields, leading to the coexistence of the SC and N states in the nanobridge at intermediate fields, as shown schematically in Fig. 4 i-l. We confirm this coexistence by performing both 3D TDGL simulations in Fig. 4h and applying our model of the effective critical field with a blue line for one leg and a red line for another leg of the nanostructure independently. For both cases, the angular dependence and the resulting "double transition" driven by the geometric anisotropy are reproduced, in excellent agreement with the experiment. This local control of the superconducting state provides new possibilities for the realisation of reconfigurable coexistence of superconducting, mixed, and normal states in superconductors controlled exclusively by geometry.

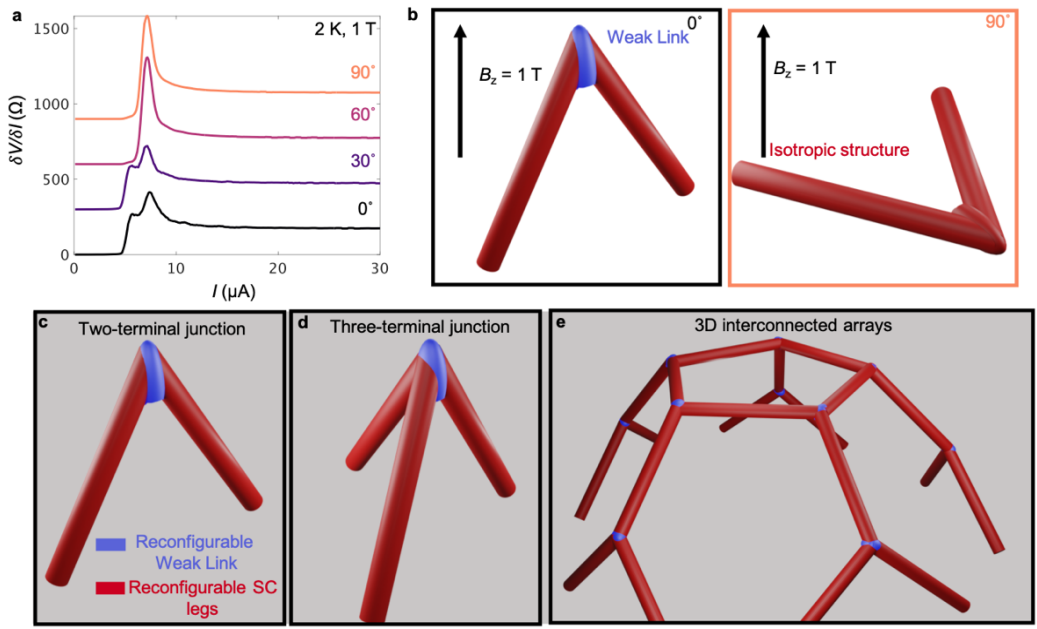

**Figure 5. Geometric anisotropy of the superconducting critical currents in the 3D nanostructure.** (a) Differential resistance as a function of applied current at several orientations of a 1 T magnetic field at 2K. (b) Sketch of the geometrical definition of a weak link, present at 0° (left), but not at 90° (right). The geometrically reconfigurable weak link offers several possibilities for devices, including (c) two-terminal and (d) multi-terminal junctions, as well as (e) a 3D interconnected network with reconfigurable weak links at the vertices.

This local control over the superconducting and normal states also opens the possibility to create Josephson weak links, which are of high interest for both fundamental and technological applications [7,9]. To determine whether we can realise geometric weak links in our 3D nanostructure, we studied the influence of the geometrical anisotropy on the critical current. In particular, we measured the current-voltage characteristic with the magnetic field rotating in the *x-z* plane, as shown in Fig. 4a. At a magnetic field of 1 T at 0° orientation, the differential resistance in Fig. 5b exhibits two pronounced peaks at $I_c \approx 5.7$ and 7.4 µA, indicating two critical currents, and therefore the presence of a 'poor' superconducting region with a lower critical current in the nanostructure that effectively functions as a weak link [54–56]. Typically, weak links are predefined by the geometry of a system, for example by the introduction of geometric constrictions or material heterostructures, meaning that the weak link is always present. Remarkably, in this 3D nanostructure, as the field is rotated to 90° the lower peak corresponding to the weak link disappears, leaving a single critical current for the entire nanostructure, indicating the presence of a geometrically reconfigurable weak link, similar to gate-tunable Josephson junctions[57].

The mechanism behind this weak link can be understood by considering the geometrical anisotropy: for all angles, both legs are symmetric with respect to the magnetic field and therefore a single transition of the legs occurs, as shown in the magnetotransport data in Fig. 4c, e. However, when the magnetic field is at 0°, at the interface between the legs – i.e. the tip of the conduit – the component of the magnetic field orthogonal to the surface is higher than the rest of the structure, meaning that in this localised region, both the critical field and the critical current are lower, resulting in a weak link. In contrast, when the magnetic field is at 90°, the component of the magnetic field orthogonal to the surface is isotropic along the structure, and the weak link disappears.

In this study, we investigate the fundamental properties of the intermediate regime of superconductivity in a single 3D nanobridge, which allowed us to demonstrate local reconfigurable control of the superconducting state and the existence of geometrically defined weak links (Figure 5c). However, for future applications the flexibility of the 3D nanoprinting allows one to extend this principle beyond simple devices, to multi-terminal 3D junctions[58] (Figure 5d) or even more complicated interconnected arrays of reconfigurable weak links (Figure 5e), thus opening the door to 3D superconducting quantum devices that go beyond the capabilities of planar systems.

In conclusion, we have realised a 3D Type II superconducting nanoarchitecture, where a 3D bridge-like geometry strongly influences the superconducting behaviour. We demonstrate the existence and propagation of superconducting vortices through the 3D conduit, paving the way to 3D fluxonic devices. By analysing the 3D configuration of superconducting vortices with the support of 3D TDGL simulations, we determine that the vortices differ from well-established Abrikosov tubes or pancakes, and instead are curved due to the geometrical confinement. This 3D geometric confinement in turn leads to a strong anisotropy in the critical field, allowing for the reconfigurable coexistence of superconducting and normal states in the nanostructure. Moreover, we demonstrate that by geometrically induced anisotropy, we can create and control weak links in the nanostructure.

With these results demonstrating the effect of geometrical confinement in 3D, we envisage a number of opportunities for the extension of superconducting nanostructures to the third dimension to impact both fundamental and applied science.

Non-standard computing architectures, such as neuromorphic and quantum computing, stand to benefit from the increase in density, complexity and interconnectivity associated with the extension to 3D superconducting nanoarchitectures. The propagation of vortices in 3D space establishes a route to 3D fluxonic applications, where again the prospects of higher density and interconnectivity offer higher complexity, with implications for 3D sensing technologies.

Finally, the geometric control of the superconducting state and the local definition of weak links, opens the door to realising reconfigurable mixed states on the nanoscale. In planar systems, such a coexistence of states has led to the realisation of Josephson junctions and a multitude of applications, including sensors and quantum computing networks. However, the configuration is typically pre-determined by the device design, geometry, and material. Here, the extension to three dimensions provides a simple route to the reconfigurable coexistence of states. In this way, it may be possible to achieve multi-functional single devices, such as sensors, logic and computing devices, or neuromorphic networks [21,59].

**Acknowledgments:**

E.Z. L.T. and C.D. acknowledge funding from the Max Planck Society Lise Meitner Excellence Program and ERC Starting Grant 3DNANOQUANT 101116043. AFP acknowledges funding from the European Research Council (ERC) under the European Union's Horizon 2020 research and innovation programme, grant agreement no. 101001290 (3DNANOMAG). V.M.F. gratefully acknowledges support from the European Cooperation in Science and Technology COST Action CA21144 (SuperQuMap) and ZIH TU Dresden as well as fruitful discussions with R. Córdoba and O. V. Dobrovolskiy.

**Author contributions:**

CD conceived the project, and EZ and CD designed the experiment. EZ conducted experimental measurements. Data analysis was performed by EZ, LT and CD. Conceptual analysis was performed by EZ, LT, U.V., VF and CD. EZ fabricated samples with the support of MK, AFP and CD. LT, WX, DS, CA, and VF performed theoretical calculations. PS and WCC performed transmission electron microscopy. EZ and CD wrote the manuscript with input from all authors.

**Competing interests:**

Authors declare no competing interests.

**Data availability:**

All data associated with this manuscript will be made available on a repository when the manuscript is published.

**Code availability Statement:**

All analysis code and reconstruction algorithms associated with the work of this manuscript will be made available on a repository when the manuscript is published.

**Methods / Supporting Information:**

*3D FEBID WC*

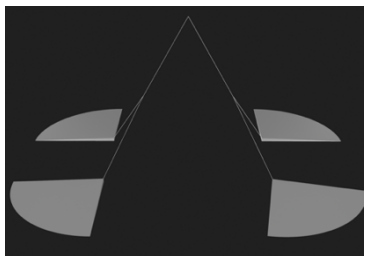

Figure S1. CAD model of the 3D nanobridge

We deposited the 3D nanostructures presented in the manuscript using a FEI Helios NanoLab 600i DualBeam microscope and f3ast software: FEBID 3D Algorithm for Stream File generation [34] to fabricate a 3D nanobridge from CAD model shown in Fig. S1. To print the 3D nanobridge, we used the electron column of the microscope, in combination with $W(CO)_6$, a standard tungsten precursor at room temperature. Deposition parameters for the focused electron beam are: $I = 11$ nA, $V = 30$ kV. The gas flow was set to 1% of the maximum available, using a commercial MultiChem Gas Injector System.

Unlike deposition with ion beams, the f3ast software for FEBID allows the nanostructure fabrication from the CAD file, opening a door for 3D superconducting nanostructures with arbitrary geometries.

*Transmission electron microscopy of 3D FEBID WC:*

We performed a structural investigation of the 3D FEBID structures via transmission electron microscope (TEM) for a deeper understanding of superconductivity in the 3D nanostructures. In Fig. S2, we show an exemplary TEM micrograph of the tip of the nano-structure with the same geometry and deposition parameters as the sample in the main text. In the TEM micrograph, the structure looks homogeneous and highly disordered. The corresponding fast Fourier transform (FFT) in Fig. S2 shows a characteristic ring with a diameter of 2.29 Å (see white labels in Fig. S2 right) and points corresponding to amorphous or at last highly polycrystalline tungsten. Additionally, we find discrete reflections of the tungsten-carbide phase (see yellow labels in Fig. S2 right), suggesting the nanocrystalline or nanocomposite structure of the nanobridge. The EDX (energy dispersive X-ray spectroscopy) analysis (spectra not shown) shows atomic concentrations up to 85 % W, 8 % C, and 7 % O in the tip of the 3D FEBID nanobridge.

TEM analysis was carried out using a FEI Tecnai F30-G2 with Super-Twin lens (FEI) with a field-emission gun at an acceleration voltage of 300 kV. The point resolution amounted to 2 Å, and the information limit amounted to about 1.2 Å. The microscope is equipped with a slow scan CCD camera (MultiScan, 2k × 2k pixels; Gatan Inc., Pleasanton, CA, USA).

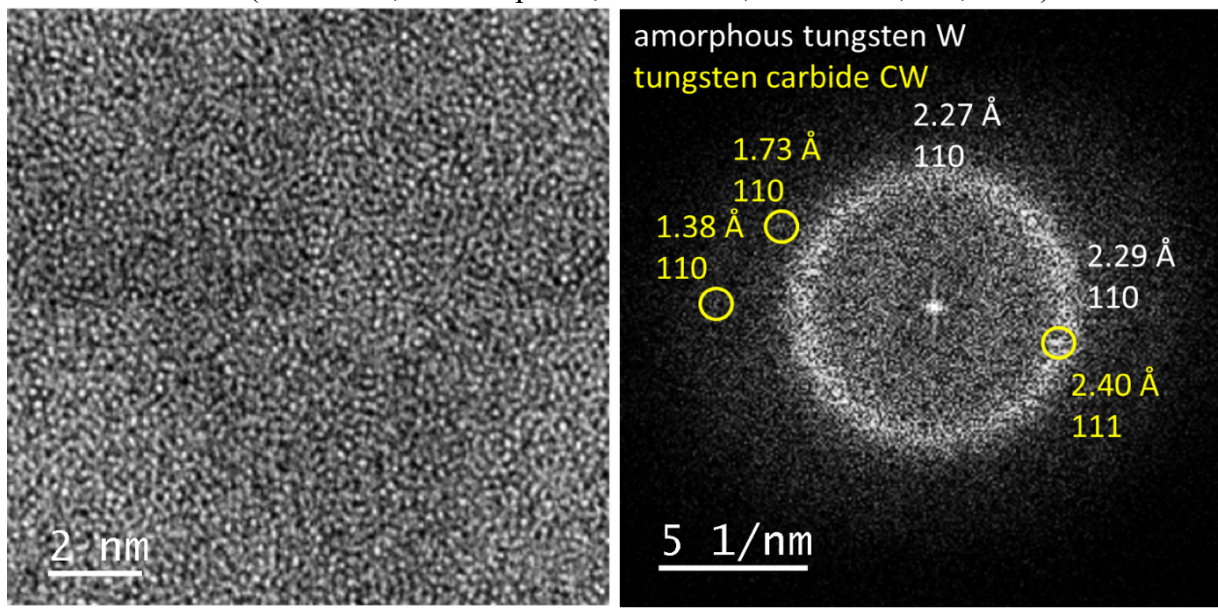

Figure S2 (Left) High-resolution TEM micrograph of the 3D FEBID nanostructure written with the deposition parameters $V$ = 30 kV, $I$ = 11 nA and gas flow of 1 % and (right) corresponding FFT.

*Resistivity measurements and estimate of SC parameters*

During focused electron beam-induced deposition (FEBID) or focused ion beam-induced deposition (FIBID) on the surface of the substrate, carbon compounds are formed by the decomposition of the organometallic gas around the desired structure, which can extend to several micrometers. This 'parasitic' deposition is crucial for electrical transport measurements. The extra deposited material between contacts electrically shorts them, causing incorrect resistance measurements. Therefore, we advanced the 3D deposition procedure for extremely high-precision electrical transport measurements in the following way.

We used a Silson silicon nitride membrane of 150 nm thickness as a primary substrate. Then, the membrane was sputtered with a 100 nm gold layer to ensure good electrical contact. To avoid an electrical shortage of the contacts, we cut a cross of around 200 to 500 nm width in a membrane, as shown in Fig. S3 to isolate four contacts from each other electrically. In this geometry, there is an empty space between the contact, therefore, no 'parasite' redeposition of material is possible. After preparing the substrate, we write a 3D nanostructure directly on the gold contacts. To reduce contact resistance, we introduced wide contact pads into the 3D geometry. This vertex structure consists of two nanowires that meet in the center and is grown in a single deposition on four legs directly on isolated contacts, allowing for 4-probe electrical transport measurements as shown in Fig. S3a.

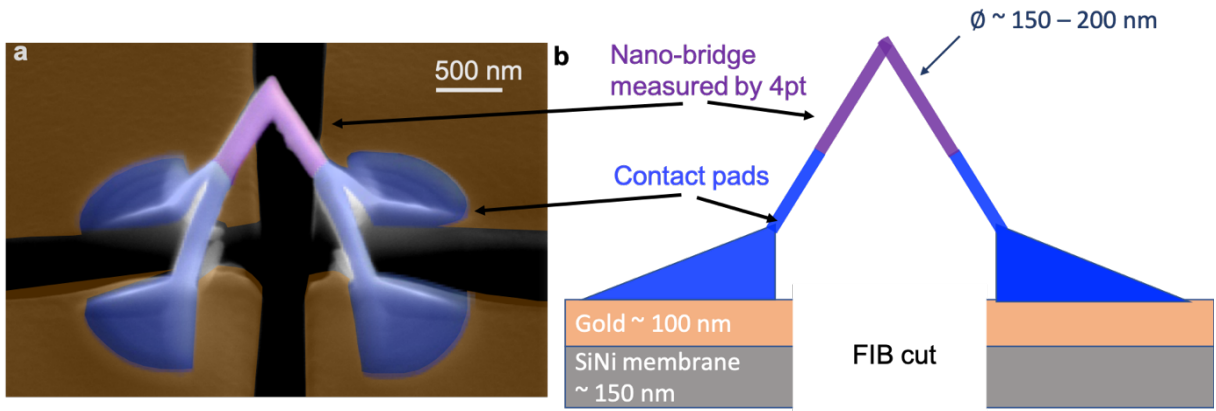

Figure S3 (a) Colored SEM image of the 3D FEBID nanobridge showing contacts and geometry of the four-point resistivity probe measurements. (b) Side view sketch of the 3D FEBID nanobridge on the membrane.

The procedure outlined herein effectively eliminates the residual 'parasite' resistance, significantly increasing experimental data quality. All electrical transport measurements were made using a bespoke low-noise probe placed within a Quantum Design Physical Property Measurement System. Standard alternating current lock-in techniques were used for the voltage measurements at a frequency of 123.3 Hz and with currents up to 30 µA, using a dual-end current source providing high common mode rejection and a Synktek MCL1-540 multichannel lock-in provides a noise level of less than 1.8 nV/√Hz. The magnetic field in the magnetoresistance measurements ranged between –9 and +9 T.

Conducting these measurements, we estimated one of the most vital superconducting parameters, which is the critical temperature (at which the SC-N transition occurs) for the 3D FEBID nano-bridge. The nano-bridge exhibits a distinct resistance drop at SC-N transition at Tc around 5 K (the value measured at 0.5 of normal state resistance $R_N$) (as illustrated in Figure 1c). Importantly, our results exhibit high consistency, reliability, and reproducibility as multiple nanostructures with similar geometries and identical deposition parameters consistently demonstrate critical temperature values within the 4.8 to 5.1 K range.

In order to study the influence of 3D geometry on superconducting properties, we characterise the dependence of the upper critical magnetic field $\mu_0 H_{c2}(T)$ as a function of temperature in Fig. 1d, which for Type II three-dimensional bulk samples or two-dimensional thin films are expected to follow this power law $\mu_0 H_{c2}(T) \sim (1 - T/T_c)^n$ with $n$ close to 1, while a value of $n \sim 0.5$ is predicted for one-dimensional wires [60,61]. For our system, we find that the power law is indeed followed, confirming the Type II behaviour, however, our nanobridge exhibits an intermediate value of $n = 0.79 \pm 0.04$, suggesting an intermediate regime of dimensionality due to the nanoscale confinement, consistent with helium-ion beam deposited structures [27,44].

Type II superconductors exhibit two fundamental length scales: the Cooper pair coherence length ξ, which represents the distance over which, in phenomenological Ginsburg- Landau theory, the superconducting order parameter gradually changes from its maximum value to zero, and penetration length λ that characterises the depth over which an external magnetic field can penetrate a superconducting material. By measuring $\mu_0 H_{c2}(T)$ as a function of temperature, we deduce the Cooper pair coherence length using the orbital limit estimation of $\mu_0 H_{c2}(T) = \Phi_0/2\pi\xi^2(T)$ (with the experimental value of $\mu_0 H_{c2}(0 \text{ K}) = 9.7 \pm 0.6$ T) to be ξ = 5.75 ± 0.15 nm. Transmission electron microscopy reveals that FEBID tungsten-carbide exhibits a highly disordered homogenous nano-crystalline nature, with a typical size of the disorder less than 3-5 nm, which is in agreement with previous studies of the tungsten-carbide with a similar deposition method [62,63]. Therefore, having a disorder smaller than ξ in the material allows us to consider the nanobridge as a type II superconductor in a weak coupling regime and in the dirty limit. The magnetic field penetration length λ can be estimated by applying the expression from the Gor'kov theory $\lambda(0) = 1.05 \times 10^{-3}\sqrt{\rho_N/T_c}$ [64]. Consequently, λ is found for several structures to vary from 850 ± 25 nm to 1.2 ± 0.1 μm. Therefore, the diameter ($D$) of the nanobridge leg satisfies the relationship ξ ≪ $D$ ≪ λ, consistent with helium-ion beam deposited structures [27,44].

*Local magnetoresistance measurements*

We performed standard local measurements to better understand the flow regime in the nano-bridge along with the non-local measurements presented in the main text. These schematics are shown in Fig. S4. In this geometry, the current applied across red contacts passes through the conduit of the nanobridge (grey) along the *y*-axis, and voltage is measured along this direction (blue contacts) with the field applied along *z*-axis orthogonally to the substrate. The magnetoresistance of the sample is shown in Fig. S4b in a temperature range from 2 to 6 K. With the increase of the temperature, the critical field at which the nanobridge exhibits SC-N transition decreases, and at 6 K the superconductor transition disappears, leaving the almost field-independent normal state magnetoresistance. The values of the depinning $I_d$ current are determined from the non-local measurements performed in the same sample.

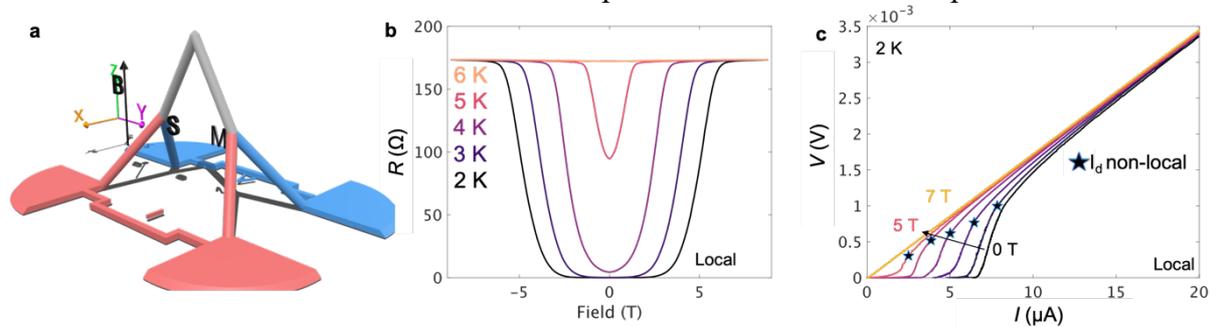

Figure S4 (a) sketch of the local measurement (b) Magnetoresistance as a function of an applied magnetic field at a temperature range from 2 to 6 K. (c) Voltage-vs-Current characteristics of the nanostructure in a transverse geometry under a perpendicular magnetic field, from 0 T to 5 T and 7 T. $I_d$ is the depinning current determined from the non-local measurements.

The disorder-driven pinning is related to the microstructure of our material [51]. To determine the microstructure of our FEBID tungsten carbide, we perform transmission electron microscopy, which reveals that the deposited material is isotropic, consisting of nanocrystallites with an average size of 3-5 nm, smaller than the size of the vortices (~20 nm), which suggests a low disorder-driven pinning regime [40,65,66]. The homogeneous nature of our superconducting nanostructure is further confirmed by the observed sharp drop in resistance at the SC transition in Fig.1c (without the presence of any resistive tail) [67,68].

*Vortex propagation in 3D nanostructure*

In the non-local measurement configuration shown in Fig. S5a a magnetic field is applied perpendicular to both the substrate plane the 3D conduit region, as illustrated in Figure 2a. An electrical current is then applied between two neighbouring contact legs (red) orthogonally to the primary channel of the conduit (grey), such that the current does not pass through the conduit (grey), and a non-local voltage is measured between the opposite set of contact legs (blue). In this geometry of electrical contacts, in the absence of the potential difference across the blue voltage contacts ($V_{nl}$), the magnetoresistance ($R_{nl} = V_{nl}/I$) is negligibly small, which we see for zero applied magnetic field ($B_z$) in Fig. S5b.

When $B_z > H_{c1} \sim 1/\sqrt{2}\kappa \, H_{c2} \sim 250$ mT, vortices are subjected to a local Lorentz force $F_L \sim I_x * B_z$, that acts along the y-axis. The component of the Lorentz force along the conduit $\sim F_L*\sin(\beta)$ (where $\beta$ is determined by the geometry) acts on the vortices (see Fig. S5 a). At the same time, due to defects, impurities, or irregularities of the nanostructure, vortices are subjected to the pinning force $F_p$ hindering the movement of vortices (see Fig. S5a), therefore no magnetoresistance is observed at low fields in the 'green' region in Fig. S5b. With the increase of $B_z$ the applied Lorentz force becomes higher than $F_p$ inducing the vortex motion in the nanostructure as shown in Fig. S5a. When vortices move they create an electric field $\boldsymbol{e} = -\boldsymbol{v_{Ly}} \times \boldsymbol{\nabla}(\frac{m^*v_s}{e^*})$ (where $v_s$ is the vortex circulation velocity, $m^*$ and $e^*$ is an electron mass and charge) schematically demonstrated in Fig. S5a [69,70]. As the conduit is populated by vortices, momentum is transferred to other vortices along the grey conduit (Fig. S5a) via repulsive vortex-vortex interaction and is supported by edge confinement along that section. The average electric field $\boldsymbol{E} = \boldsymbol{B} \times \boldsymbol{v}_{Ly}$ [40,42,70] associated with a moving vortex in the vicinity of the voltage probe generates a potential difference and correspondingly the non-local voltage, which results in non-zero magnetoresistance in the orange region in Fig. S5b.

We note that standard alternating current lock-in techniques were used for the voltage measurements at a frequency of 123.3 Hz with a current magnitude of 5 µA. As a result, an alternating Lorentz force was felt locally by the vortices, corresponding to a "push-pull" effect. Through the measurement of the low-frequency first harmonic response presented in Figure 2, we therefore detect the non-local motion of vortices.

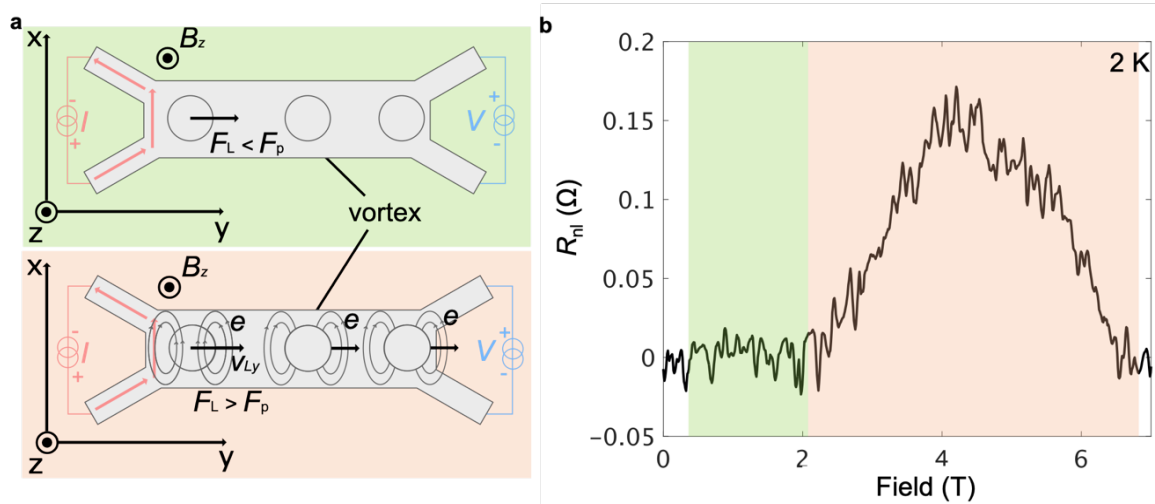

Figure S5. (a)Schematics of the vortex motion under the application of Lorentz force with corresponding (b) magnetoresistance measurements in the non-local geometry at 2 K and AC current $I = 5$ µA.

*Anisotropy effects*

Considering the surface effect discussed in the main text in comparison to the anisotropy in 2D thin films where in the parallel orientation of the film to the magnetic field, the dimension of the current loops is fixed by the thickness, hence is constant, on the other hand, in the perpendicular orientation, the size of the current loops scales down as magnetic field increases [53]. Our 3D nanostructure is in the intermediate regime between these two limits, as the geometrical restriction of the superconducting current loops is always present and depends on the orientation of the magnetic field and, therefore, is determined by the component orthogonal to the surface.

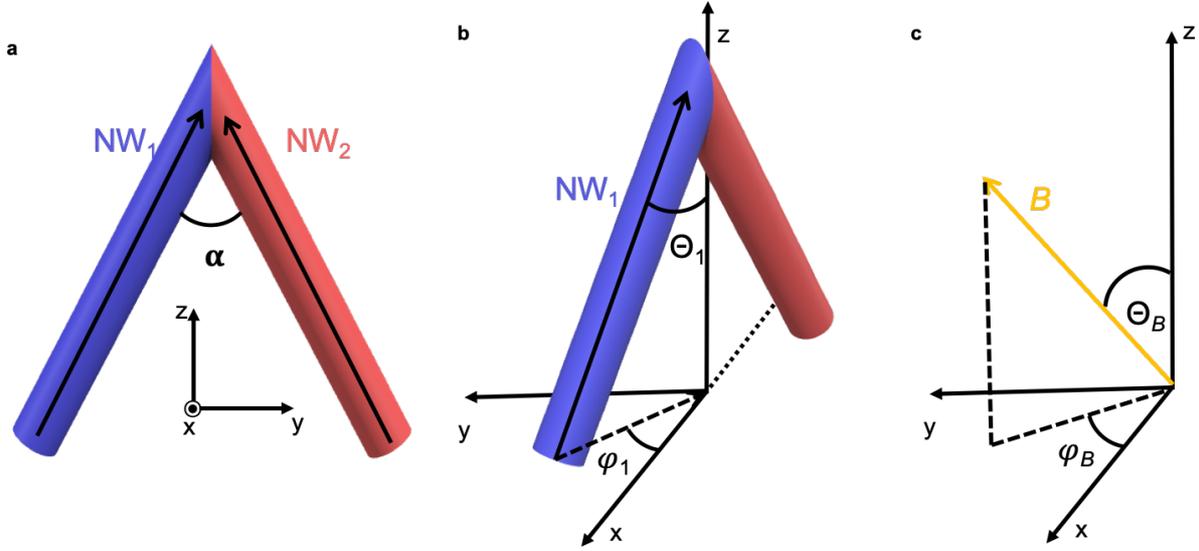

Figure S6 a) and b) sketch showing the angles of the two vectors **NW₁** and **NW₂** corresponding to two legs forming the nanostructure. c) Rotation of the applied magnetic field

To calculate this component in 3D, we first need to set up the coordinates of the structure and magnetic field. We will consider each leg of the nanobridge independently and determine the coordinates of the vectors along the long axis of the nanowires. Therefore, as shown in Fig. S6a and b, we have two vectors **NW₁** = $(\cos(\Theta_1)\sin(\varphi_1), \sin(\Theta_1)\sin(\varphi_1), \cos(\Theta_1))$ and **NW₂** = $(\cos(\Theta_2)\sin(\varphi_2), \sin(\Theta_2)\sin(\varphi_2), \cos(\Theta_2))$, where $\Theta_1$ and $\Theta_2$ are the inclination angles of the legs ($\Theta_1 + \Theta_2 = \alpha$, $\alpha$ is angle between the legs, set up by the geometry of the structure) $\varphi_1$ and $\varphi_2$ determine the plane of the conduit. In this model, we define the two vectors of the legs $NW_1$ and $NW_2$ and maintain them unaltered. In the next step, we define the rotation of the magnetic field. The vector of the magnetic field **B** = $(\cos(\Theta_B)\sin(\varphi_B), \sin(\Theta_B)\sin(\varphi_B), \cos(\Theta_B))$, where $\varphi_B$ defines the plane in which the magnetic field is rotated, and $\Theta_B$ defined the angle of rotation of the magnetic field (see Fig. S6c).

In this way, we can model this anisotropy as a correction to the minimal critical field $H_c = H_{cb} + \Delta$, as the introduction of the curvature to the vortices increases the upper critical field [71], where $H_{cb}$ is a constant critical field determined from the experiment as the minimal critical field, and $\Delta$ is a calculated correction $\sim \gamma/H_{perp} \sim \gamma/|\sin(\eta)|$, where $\gamma$ is a fitting parameter and $\eta$ is an angle between the magnetic field **B** and the vectors of the legs **NW₁** or **NW₂**.

As discussed in the main text, first, we perform measurements with the rotation of the magnetic field in the *z-x* plane (perpendicular to the plane of the conduit). Therefore, we can define the magnetic field vector by $\varphi_B = 90°$, and $\Theta_B$ is varied from 0 to 180°. The measurement of the superconducting transition in this orientation of magnetic field rotation in Fig. S7a is not ideally symmetric and shows a splitting of the transition at low rotation angles, indicating the slight initial misalignment of the plane of the conduit leading to inequivalence of two legs. Therefore, we can estimate **NW**$_1$ by $\Theta_1 = 21°$ (the inclination of the leg from the vertical line in Fig. S6a), $\varphi_1 = 15°$ (the misalignment of the plane of the conduit) and **NW**$_2$ by $\Theta_1 = -39°$ (the inclination of the leg from the vertical line in Fig. S6a), $\varphi_1 = -15°$ (the misalignment of the plane of the conduit). The result of the calculation of the critical magnetic field with this model is shown in Fig. S7a on top of the experimental results. For this field rotation, the fitting parameters $H_{cb} = 2$ T, $\gamma_1 = 0.35$, $\gamma_2 = 0.7$.

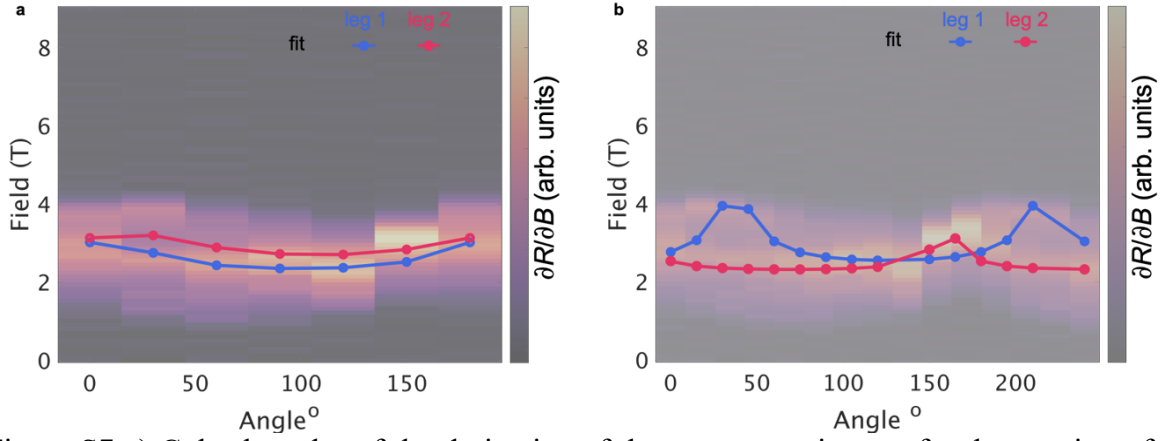

Figure S7 a) Color bar plot of the derivative of the magnetoresistance for the rotation of the magnetic field in the *z-x* plane. The model calculated independently for both legs of the conduit is shown on top of the experimental data. b) the same for the rotation of the magnetic field in the *y-z* plane

Next, we rotate the magnetic field in the *y-z* plane (in the plane of the conduit) so that the legs of the conduit are aligned at different angles to the magnetic field. Now we can define the magnetic field vector by $\varphi_B = 0°$, and $\Theta_B$ is varied from 0 to 240°. The vectors of both legs: **NW**$_1$ with $\Theta_1 = 39°$ (the inclination of the leg from the vertical line in fig. 1a), $\varphi_1 = 15°$ (the misalignment of the plane of the conduit) and **NW**$_2$ with $\Theta_1 = -21°$ (the inclination of the leg from the vertical line in Fig. S6a), $\varphi_1 = -15°$ (the misalignment of the plane of the conduit). The measurement of the superconducting transition in this orientation of magnetic field rotation is in Fig. S7b shows the split transition, indicating that the two legs of the conduit undergo a superconducting to normal metal (SC-N) transition at different magnetic fields, leading to the coexistence of the SC-N state in the nanobridge at intermediate fields. The calculated model (with fitting parameters $\gamma_1 = 0.35$, $\gamma_2 = 0.12$) for both legs is on top of the experimental data in Fig. S7b and reproduces both the anisotropy and the resulting 'double transition'.

*3D TDGL*
We performed time-dependent Ginzburg Landau (TDGL) simulations to determine the superconducting vortex structure in the three-dimensional nanostructures. The TDGL equations relate the superconducting order parameter $\Psi$ to the magnetic vector potential **A** and the superconducting current **J**$_s$, as a function of time. Here, we use unitless TDGL equations to establish the equilibrium configurations of the superconducting vortices. The equations take the following form:

$$\eta \frac{\partial \Psi}{\partial t} = -\left(\frac{i}{\kappa}\vec{\nabla} + \vec{A}\right)^2 \Psi + [\varepsilon - |\Psi|]^2 \Psi,$$

$$\sigma \frac{\partial \vec{A}}{\partial t} = \vec{J}_s - \vec{\nabla} \times \vec{\nabla} \times \vec{A},$$

where $\eta$ is the ratio of the characteristic times for the relaxation of the order parameter and the vector potential, $\kappa$ is the ratio of the penetration depth and the superconducting coherence length, and $\sigma$ is the conductivity of normal state electrons [72]. $\varepsilon$ is the pinning coefficient, and is set to 1 (full superconductivity). The superconducting current is given by

$$\vec{J}_s = \frac{1}{2\kappa i}(\Psi^*\vec{\nabla}\Psi - \Psi\vec{\nabla}\Psi^*) - |\Psi|^2\vec{A}.$$

The value of $\kappa$ was taken to be 140, as determined experimentally, and as we are only interested in the equilibrium solutions of the TDGL equations, $\eta$ and $\varepsilon$ were set to 1.

The simulations were performed using a finite element implementation of TDGL theory within the COMSOL Multiphysics software package [73]. The coupled equations were represented as a single, general form partial differential equation

$$d\frac{\partial \vec{u}}{\partial t} + \vec{\nabla} \cdot \vec{\Gamma} = \vec{F}$$

akin to the approach of Oripov et al. [74]. We implemented a two-domain approach, simulating a superconducting cylindrical nanowire, embedded within a spherical vacuum. To preclude any current passing through the boundary of the nanowire, $\partial\Omega$, to the vacuum, we imposed the boundary condition

$$\vec{J}_s \cdot \hat{n} = 0 \text{ on } \partial\Omega$$

where $\hat{n}$ is the surface normal of the boundary [75]. To ensure the continuity of the magnetic field across the nanostructure boundary, we also impose

$$\vec{\nabla} \times \vec{A} = \vec{\nabla} \times \vec{A}_{ext} \text{ on } \partial\Omega.$$

The Dirichlet condition that the vector potential must equal the vector potential of the external magnetic field was applied to the boundary of the vacuum:

$$\vec{A} = \vec{A}_{ext} \text{ on } \partial\Omega_{vac}$$